\documentclass[twocolumn,prx,superscriptaddress]{revtex4-2}
\usepackage{amsmath,amssymb,mathrsfs}
\usepackage{natbib}
\usepackage{subfigure}
\usepackage{tabularx}
\usepackage{epsfig}
\usepackage{longtable}
\usepackage{amsfonts}
\usepackage{rotating}
\usepackage{bbold}
\usepackage{hhline}
\usepackage{braket}
\usepackage{txfonts, comment} 
\usepackage{bbm} 

\usepackage[unicode=true,bookmarks=true,bookmarksnumbered=false,bookmarksopen=false,breaklinks=false,pdfborder={0 0 1},backref=false,colorlinks=true]{hyperref}

\hypersetup{linkcolor=magenta,urlcolor=blue,citecolor=blue,pdfstartview={FitH},hyperfootnotes=false,unicode=true}

\def\blue{\color{blue}}

\def\be{\begin{equation}}
\def\ee{\end{equation}}
\def\bea{\begin{eqnarray}}
\def\eea{\end{eqnarray}}

\newcommand{\norm}[1]{\left\lVert#1\right\rVert}

\begin{document}
\title{Configured Quantum Reservoir Computing for Multi-Task Machine  Learning}

\author{Wei Xia}
\affiliation{State Key Laboratory of Surface Physics, Key Laboratory of Micro and Nano Photonic Structures (MOE), and Department of Physics, Fudan University, Shanghai 200433, China}

\author{Jie Zou}
\affiliation{State Key Laboratory of Surface Physics, Key Laboratory of Micro and Nano Photonic Structures (MOE), and Department of Physics, Fudan University, Shanghai 200433, China} 

\author{Xingze Qiu}
\affiliation{State Key Laboratory of Surface Physics, Key Laboratory of Micro and Nano Photonic Structures (MOE), and Department of Physics, Fudan University, Shanghai 200433, China}
\affiliation{School of Physics Science and Engineering, Tongji University} 
\author{Feng Chen} 
\affiliation{Institute of Science and Technology for Brain-Inspired Intelligence, Fudan University, Shanghai 200433, China}
\author{Bing Zhu} 
\affiliation{HSBC Lab, HSBC Holdings plc, Guangzhou, China} 
\author{Chunhe Li}
\affiliation{Institute of Science and Technology for Brain-Inspired Intelligence, Fudan University, Shanghai 200433, China}
\affiliation{Shanghai Center for Mathematical Sciences and School of Mathematical Sciences, Fudan University, Shanghai 200433, China}
\author{Dong-Ling Deng} 
\affiliation{Center for Quantum Information, IIIS, Tsinghua University, Beijing 100084, China} 
\affiliation{Shanghai Qi Zhi Institute, AI Tower, Xuhui District, Shanghai 200232, China}
\author{Xiaopeng Li}
\email{xiaopeng\underline{ }li@fudan.edu.cn}
\affiliation{State Key Laboratory of Surface Physics, Key Laboratory of Micro and Nano Photonic Structures (MOE), and Department of Physics, Fudan University, Shanghai 200433, China} 
\affiliation{Shanghai Qi Zhi Institute, AI Tower, Xuhui District, Shanghai 200232, China}
\affiliation{Shanghai Research Center for Quantum Sciences, Shanghai 201315, China}

\begin{abstract}
Amidst the rapid advancements in experimental technology, noise-intermediate-scale quantum (NISQ) devices have become increasingly programmable, offering versatile opportunities to leverage quantum computational advantage. 
Here we explore the intricate dynamics of programmable NISQ devices for quantum reservoir computing. 
 Using a genetic algorithm to configure the quantum reservoir dynamics, we systematically enhance the learning performance. Remarkably, a single configured quantum reservoir can simultaneously learn multiple tasks, including a synthetic oscillatory network of transcriptional regulators, chaotic motifs in gene regulatory networks, and the fractional-order Chua's circuit. 
 Our configured quantum reservoir computing yields highly precise predictions for these learning tasks, outperforming classical reservoir computing. We also test the configured quantum reservoir computing in foreign exchange (FX) market applications and demonstrate its capability to capture the stochastic evolution of the exchange rates with significantly greater accuracy than classical reservoir computing approaches. 
 Through comparison with classical reservoir computing, we highlight the unique role of quantum coherence in the quantum reservoir, which underpins its exceptional learning performance. Our findings suggest the exciting potential of configured quantum reservoir computing for exploiting the quantum computation power of NISQ devices in developing artificial general intelligence.  

%To take full advantage of the computational power of those NISQ devices, we study the potential quantum advantage of quantum reservoir computing and the learning power of quantum reservoirs trained by gene algorithm (GA). Here we construct the classical and quantum reservoirs and contrast their computational capacities to demonstrate the potential quantum advantage. Furthermore, we propose that the learning power of the quantum reservoir trained by GA can be systematically improved. The trained quantum reservoir can simultaneously learn different tasks: a synthetic oscillatory network of transcriptional regulators, chaotic motifs in gene regulatory networks, and fractional-order Chua’s circuit with a memristor. Except for those artificial systems, we also investigate the real-world problem, the exchange rate prediction, with the trained quantum reservoir. The trained quantum reservoir has acquired outstanding performance on the above tasks. Therefore, the quantum reservoir should be engineered at the optimal parameters obtained by GA to have the optimal learning power for generic tasks.

\end{abstract}

\date{\today}
\maketitle

%{\it Introduction.---}
\section{Introduction}
The past two decades have witnessed rapid developments in quantum technologies. 
The advancements in quantum computation have been particularly impressive, with demonstrations of quantum advantage on certain tasks using NISQ devices, such as random circuit sampling~\cite{2019_Google_Nature} and boson sampling~\cite{2020_Pan_Science}.
The search for practical applications with NISQ devices~\cite{Preskill2018quantumcomputingin} has received immense research efforts, leading to the creation of several quantum computing approaches such as the quantum approximate optimization algorithm~\cite{farhi2014quantum,harrigan2021quantum,ebadi2022quantum}, the variational quantum eigensolver~\cite{ebadi2022quantum,peruzzo2014variational,kandala2017hardware}, and adiabatic quantum computation~\cite{RevModPhys.90.015002,hauke2020perspectives,farhi2000quantum,farhi2001quantum}. Recently, a novel computation framework, known as quantum reservoir computing (QRC)~\cite{Keisuke2017}, has emerged. 
The QRC framework accomplishes machine learning tasks by mapping the input signal to a high-dimensional space having complex quantum superposition, which connects to the desired output through a linear regression model or a relatively simple neural network. 
This approach to harnessing the power of quantum computation has attracted rapidly growing attention due to its unique experimental accessibility to NISQ devices~\cite{xia2022reservoir,negoro2018machine,chen2020temporal,dasgupta2022characterizing,bravo2022quantum}.

%ghosh2021quantum,mujal2021opportunities}. 
%In addition to these approaches, a new computation framework, quantum reservoir computing (QRC)~\cite{Keisuke2017}, has recently gained significant attention due to its low training cost and potential for hardware implementation~\cite{ghosh2021quantum,mujal2021opportunities}. 
%The QRC system consists of a quantum reservoir that maps inputs into a high-dimensional space and a single linear-layer neural network for readout. The quantum reservoir is fixed, and only the readout layer is trained, which makes it advantageous for experimental realization on near-term quantum devices. 
%Furthermore, the QRC framework allows the parameters of the quantum reservoir to be randomly selected within limits, and it does not require precise control of the quantum reservoir. As a result, the QRC is advantageous for experimental realization on near-term quantum devices and has the potential to solve practical problems~\cite{negoro2018machine,chen2020temporal,dasgupta2022characterizing,bravo2022quantum}. This work proposes four different reservoirs to demonstrate the potential quantum advantage of the QRC framework.

In current QRC models, the many-body Hamiltonian that governs the quantum reservoir dynamics remains fixed and untouched during the learning process. However, it has been found that different Hamiltonian constructions can lead to significantly different learning performance~\cite{Keisuke2017,xia2022reservoir, martinez2021dynamical,llodra2023benchmarking}. To maximize the learning performance, a guiding principle for constructing the Hamiltonian has been proposed, which involves engineering the reservoir dynamics near the phase boundary of quantum ergodicity~\cite{xia2022reservoir, martinez2021dynamical}. It has been found that quantum criticality enhances the QRC capability.
Despite these advances, the learning tasks achieved by QRC are still relatively restricted to simple tasks, such as parity checking, short-term memory, and small-scale NARMA tasks. 
Further  innovations are  required for the development of QRC for more complex learning tasks, and for artificial general intelligence. 
%Development of QRC for more complex learning tasks demands further innovation. 
%Thus, there is still significant room for improvement in the development of QRC for more complex learning tasks.

%For a given classical reservoir~\cite{packard1988adaptation,bertschinger2004rea,Mushegh_PRX2020,Carroll_arXiv2019,LEGENSTEIN2007323}, it has been suggested that systems with dynamics that fall somewhere between order and chaos could achieve significant computational power. This phenomenon has also been observed in QRC~\cite{xia2022reservoir, martinez2021dynamical}. The optimal learning power is achieved when approaching the boundary of quantum ergodicity. These findings suggest that by selecting the appropriate parameters within a certain range, QRC could demonstrate a higher level of computational power. These results are important for implementing QRC experiments on near-term quantum devices and maximizing their potential. However, it is worth noting that many quantum systems lack a well-defined phase transition or exact phase boundary. In such cases, it is unclear what the ideal parameters are for boosting processing power.

\begin{figure*}[htp]
%\vspace{-0.2 cm}
\includegraphics[width=.8\linewidth]{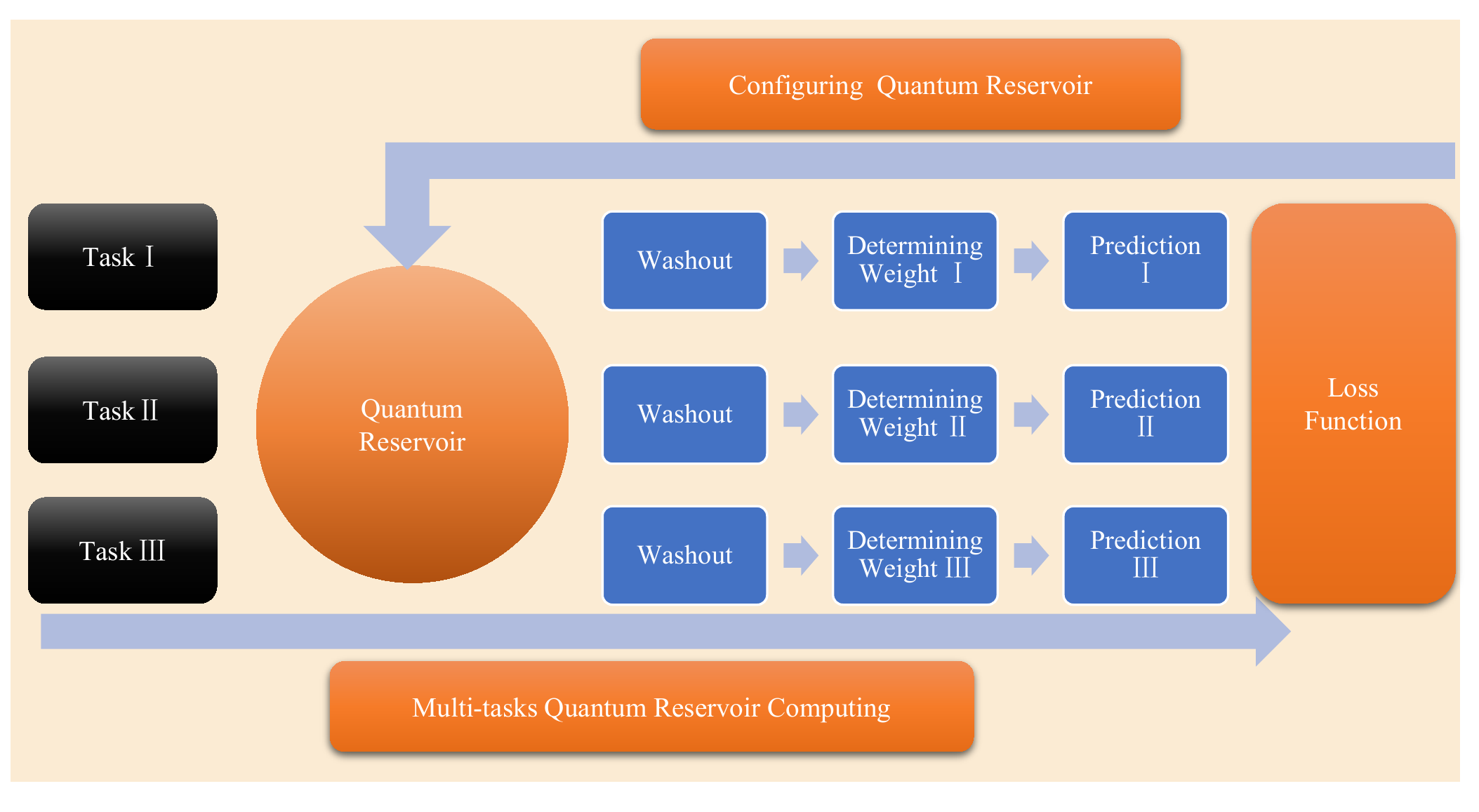}
\caption{
{\bf The schematic diagram of the configured quantum reservoir computing for multi-task machine learning.} 
The forward process denotes multi-task quantum reservoir computing. In this process, the inputs of three tasks are  sequentially injected into the quantum reservoir. The initial time evolution steps of the reservoir are used for a standard information washout. The observables of the quantum reservoir are collected and then transformed to compute the outputs by a linear regression model. The backward process represents configuring the quantum reservoir by a genetic algorithm.  During this process, the total loss function of three tasks is evaluated and the quantum reservoir is optimized accordingly.  The forward and backward processes are iterated until the loss function converges. 
}
\label{Fig1}
\end{figure*}

Here, we propose a novel approach to QRC that enables simultaneous learning of multiple complex tasks. Our approach configures the quantum reservoir dynamics to optimize performance on multiple tasks using a genetic algorithm, analogous to the biological evolution of human intelligence. Configured quantum reservoirs exhibit sufficient computational capacity to tackle real-world problems. Through numerical tests, we demonstrate that a single configured quantum reservoir can  handle multiple tasks including synthetic oscillatory networks of transcriptional regulators~\cite{elowitz2000synthetic}, chaotic motifs in gene regulatory networks~\cite{zhang2012chaotic}, and fractional-order Chua's circuits~\cite{petras2010fractional}. 
% In all cases, configured quantum reservoirs significantly outperform randomly constructed ones, which are commonly used in conventional QRC~\cite{Keisuke2017,xia2022reservoir,martinez2021dynamical,kutvonen2020optimizing}.
In all cases, the configured quantum reservoirs significantly outperform  the echo state network (ESN) method~\cite{jaeger2004harnessing}, a widely used approach in classical reservoir computing 
~\cite{tanaka2019recent,tong2007learning,skowronski2007automatic,jaeger2007optimization,Decai2012Chaotic}. 
We attribute the quantum advantage of the configured quantum reservoirs to the quantum coherence embedded within the quantum reservoir. Furthermore, we apply our approach to FX market applications, 
specifically predicting the exchange rates of GBP/USD, NZD/USD, and AUD/USD 
with significantly greater accuracy than classical reservoir computing approaches investigated in previous studies~\cite{wang2022echo}. 
This study demonstrates outstanding learning performance and the remarkable transferability of our configured quantum reservoir computing. 
Multi-task learning with configured quantum reservoirs provides a compelling computational model for establishing the quantum advantage of NISQ devices in practical applications and paves the way for further development of artificial general intelligence.  

%The design of the reservoir~\cite{chen2019learning, chen2020temporal}, encoding basis~\cite{xia2022reservoir,caro2021encoding}, and training weight~\cite{wyffels2010comparative} significantly affect its computational power. However, finding the optimal parameters through random experiments is an exhausting process. To address this issue, we propose training quantum reservoirs and systematically improving their computational power. The resulting quantum reservoirs have the sufficient computational capacity to tackle real-world problems and support multi-task learning. For example, we demonstrate the effectiveness of our approach by forecasting three higher-dimensional sequences simultaneously: a synthetic oscillatory network of transcriptional regulators~\cite{elowitz2000synthetic}, chaotic motifs in gene regulatory networks~\cite{zhang2012chaotic}~\cite{petras2010fractional}, and fractional-order Chua's circuit with a memristor. Details about these sequences are provided in the Method section. In addition, we apply our approach to a real-world problem involving exchange rates, which is a complex financial system. Specifically, we investigate three exchange rates: NZD/USD, USD/CHF, and AUD/USD. Using our trained quantum reservoir, we predict the exchange rate under a multi-task learning framework. We achieve this by using a quantum reservoir that has been trained on other exchange rates to predict the target exchange rate.

\section{The theoretical framework } 
\label{sec:theory} 

The theoretical framework of our configured quantum reservoir computing is illustrated in Fig.~\ref{Fig1}. 
The quantum reservoir dynamics is governed by a parameterized Hamiltonian $\hat{H}(\theta)$, where $\theta$ represents the controllable parameters of the reservoir. These parameters are configured to optimize the overall learning performance. The input and output setup of this quantum machine learning model remains consistent with  the conventional quantum reservoir computing~\cite{Keisuke2017}. Both the input and output are time sequences denoted as column vectors, ${\bf s}_k$ and ${\bf y}_k$, respectively, with $k\in [1,K]$ labeling the time steps. The input and output signal dimensions, 
$d_{\rm in}$ ($d_{\rm out}$), are determined by the learning task to perform.

The input signal is sequentially injected into the quantum reservoir, 
{ 
where  the injection corresponds to projective  measurements of the first few ({$\lceil d_{\rm in}/2\rceil$})    reservoir  qubits, 
} 
followed by resetting them to product states that  encode the input signal (see Methods). 
%The state of the quantum reservoir system at time $t$ is described by a density matrix $\rho(t; {\bf s}_{k'\in [1,k]}, \theta)$. 
The quantum reservoir is let evolve for a certain time duration ($\tau$) 
in between two successive injections.  
%$\rho(k\tau + \Delta t) = e^{iH(\theta)\Delta t} \rho(k\tau) e^{-iH(\theta) \Delta t} $. 
We perform a series of Pauli measurements within each time duration,  with the results stored as a column vector 
${\bf A}_k$, which depends on $\theta$ and $\{ {\bf s}_{k_<} \} \equiv \{ {\bf s}_{k'\in [1,k]}  \}$. 
These quantum measurement results are then transformed into the final output of the QRC through a linear regression model, 
${\bf y}_k = {\bf W} \cdot {\bf A}_k  + {\bf B}$, which  is matched to the learning target ${\bf y}^\star_k$. The weights (${\bf W}$ and ${\bf B} $) are determined by minimizing their difference. The quantum reservoir parameters are configured by minimizing the objective function, 
\be 
\label{eq:costfunction}
\min _\theta  \left[  \min_{{\bf W}, {\bf B}} \left( \sum_{{\bf s}_k}
\norm {{\bf W} \cdot {\bf A}_k (\{{\bf s}_{k_<}\}; \theta) +{\bf B} - {\bf y}_k^\star (\{{\bf s}_{k_<}\})  } \right) \right], 
\ee 
where $\sum_{{\bf s}_k}$ represents summing over different input sequences of the training dataset. 
The quantum reservoir configuration, namely $\min _\theta$, is performed by a classical genetic algorithm. 
The details are provided in Methods.

The configured quantum reservoir computing has a potential quantum advantage, for the computation cost of ${\bf A}_k ({\bf s}_{k_<}; \theta)$ on a classical computer scales exponentially with the number of qubits.  Configuring the quantum reservoir is to optimize its computation capability in the context of reservoir computing.

To achieve multi-task learning, we use a single quantum reservoir, i.e., task independent, and  allow the weights of the linear regression model to be task dependent, for determining the weights of the linear  model is much less costly than the quantum reservoir. The quantum reservoir is configured to optimize  the overall learning performance on multiple tasks. 
In this way, we investigate whether a single configured quantum reservoir has the capability of learning multiple tasks simultaneously.

In this work, we choose a fully connected transverse-field Ising model as our quantum reservoir, 
\be
{\hat H} = \sum_{i,j}J_{ij}\hat{\sigma}_i^X \hat{\sigma}_j^X+\sum_i h_i \hat{\sigma}_i^Z,
\label{eq:Hamiltonian}
\ee
where $\hat{\sigma}^X$ and $\hat{\sigma}^Z$ are two Pauli operators. 
The quantum reservoir parameters are then $J_{ij}$ and $h_i$, and  having $n$ number of reservoir qubits, we have a total number of $n(n+1)/2$ such parameters. 
In the numerical simulations presented in this paper, we choose $n=6$, unless specified. 
We expect the learning performance can be further improved by increasing the number of qubits, for larger number of qubits necessarily produce more complex reservoir dynamics. 

 {\begin{figure*}[htp]
\includegraphics[width=.9\linewidth]{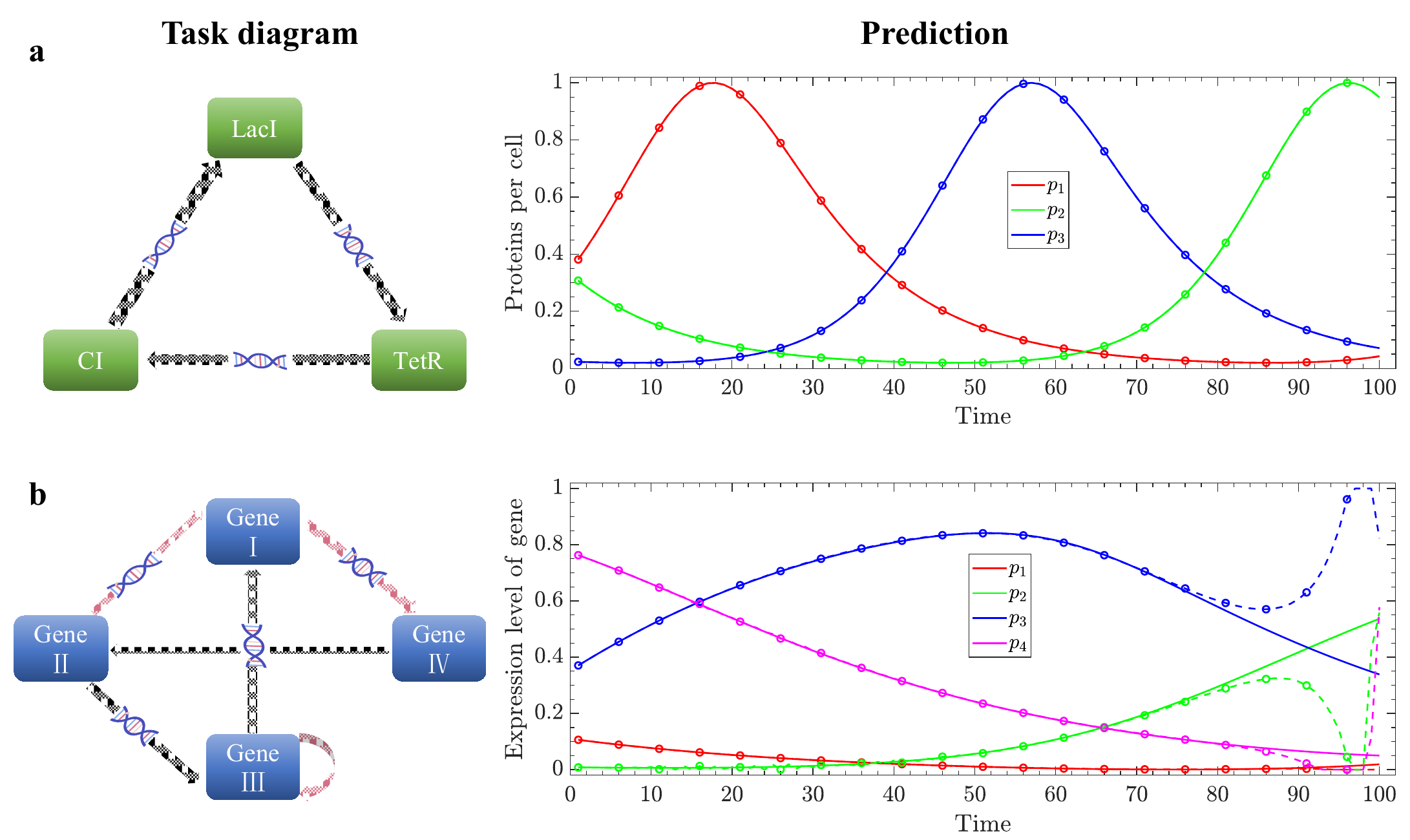}
\caption{ {\bf Application of configured quantum reservoir computing on gene regulatory networks.} {\bf a}, a synthetic oscillatory network of transcriptional regulators. The transcriptional repressors, LacI, TetR, and CI interact with each other via mutual inhibitions. Our approach correctly captures the oscillatory dynamics of the protein production, with an error unnoticeable in this plot. 
{\bf b}, a  chaotic motif gene regulatory network. This represents a biological network that can generate chaotic dynamics. 
In both {\bf a} and {\bf b}, the  inhibitory interactions are illustrated  by black arrows, and the activation interactions are illustrated by red arrows.  The solid lines and the dotted lines with circles represent the actual dynamical evolution, and the quantum reservoir prediction, respectively, in testing.
We simulate $6100$ timesteps to create  the testing dataset. 
The initial $1000$ steps are used for the reservoir washout (Methods). The first $5000$ steps are taken for determining the weights in the linear regression model. The final $100$ steps are used to test the prediction accuracy.  
}
\label{Fig2}
\end{figure*}}

\section{The performance on multi-task learning}
\label{sec:Qapp} 

{
We apply the configured quantum reservoir computing to diverse applications, including  a synthetic oscillatory network of transcriptional regulators, chaotic motifs in gene regulatory networks, fractional-order Chua’s circuits, and FX market forecast. The first three learning tasks are described by deterministic differential equations to be illustrated in detail below. The last task serves as one example having stochastic uncertainty. 

In standard machine learning applications of reservoir computing, it is typical to use the reservoir to predict the future evolution of the time sequence in the training dataset ~\cite{Keisuke2017,xia2022reservoir,martinez2021dynamical,kutvonen2020optimizing,pfeffer2022hybrid}. The time sequence  to predict thus follows the same rule as the training dataset. 
Here, we purposely make the learning more challenging to evaluate the potential of our configured quantum reservoir computing. 
The quantum reservoir is configured by optimizing the performance on the training datasets and is left untouched during testing.
To generate test datasets, we use  different parameters from those of the training dataset in the differential equations that describe the deterministic learning tasks (Methods). 
For the FX market task, we use AUD/USD and NZD/USD rates to train the quantum reservoir and test its prediction accuracy using GBP/USD.

To quantify the performance on different learning tasks, we introduce a normalized mean squared error (NMSE), 
\be 
{\rm NMSE} = \frac{\sum_k \norm{ {\bf y}_k - {\bf y}_k^\star }_2 }{\sum_k  \norm{{\bf y}_k }_2 } .
\ee
This characterizes the learning performance on different tasks at equal footing by normalization and can be used to demonstrate the advantage of our configured quantum reservoir computing over other approaches including classical reservoir computing and conventional quantum reservoir computing. 
}

%In the general learning process, the training and testing datasets are usually drawn from the same distribution. For instance, we may use $80\%$ of the data set for training and the remaining $20\%$ for testing. However, our trained quantum reservoir has extensive computational capacity and exhibits some universal ability. To demonstrate this, we design testing data that are generated from the same equations as the training data, but with different parameters. During the computation process, we train the quantum reservoir using the training data and apply the trained quantum reservoir to the testing data for prediction. Detailed parameters are provided in the Method section.

\subsection{Gene regulatory networks} 
Many complex biological processes can be modeled as dynamical systems through regulatory networks ~\cite{tsai2008robust,li2014landscape}. So, inferring the dynamical behavior of gene regulatory networks from gene expression data is vital to understanding the functions of biological systems~\cite{shen2021finding,chen2022inferring}. Here, we seek to apply quantum reservoir computing to learn the dynamics of gene regulatory networks. A representative example is the synthetic oscillatory network of transcriptional regulators, which has been proposed to model the functionality of intracellular networks~\cite{elowitz2000synthetic}. This network consists of three transcriptional repressors,  LacI, TetR, and CI, which interact with  each other through mutual inhibitions. Their interactions are illustrated in Fig.~\ref{Fig2}{\bf a}. The feedback mechanism in this network results in complex dynamics  of great interest to biological systems. 

Quantitatively, the kinetics of the synthetic oscillatory network is described by six coupled differential equations:
\be
\begin{aligned}
 \frac{dm_i}{dt} &= -m_i + \frac{\alpha}{1+p_j^{h_n}} + \alpha_0, \quad
       j = {\rm CI,LacI,TetR} \\
\frac{dp_i}{dt} &= -\beta(p_i-m_i),\quad
    i = {\rm LacI, TetR,CI}
\end{aligned}
\label{eq:Oscillatory}
\ee
Here, $p_i$ represents repressor-protein concentrations, and  $m_i$ represents corresponding mRNA concentrations. The number of protein copies per cell produced from a given promoter type is $\alpha_0$, in the presence of saturating amounts of the repressor, and $\alpha + \alpha_0$ in its absence. The protein decay rate relative to the mRNA is represented by the ratio $\beta$. The mRNA concentration is regulated by the corresponding repressor-protein, denoted by
$\alpha/(1+p_j^{h_n})$, with $h_n$ representing the Hill coefficient~\cite{elowitz2000synthetic}.

In constructing the training and test datasets, we discretize the differential equations by choosing a time step $\delta t = 0.05$. Other parameter choices are provided in Methods. 
The parameters for producing the training and test datasets are deliberately chosen to be different. With the quantum reservoir configured on the training dataset, we apply the  quantum reservoir computing to the test dataset. The weights of the linear regression model are determined by the first $6000$ steps of the time sequence.  The configured quantum reservoir computing is used to predict the $100$ forward steps. 
Its comparison with the accurate time sequence is shown in  Fig.~\ref{Fig2}{\bf a}. 
The prediction for the protein concentration correctly captures the mutual inhibition and the resultant oscillatory behavior of the network model, with the discrepancy barely noticeable.  
The maximum NMSE for this task is at the level of $10^{-10}$.

\medskip

The second gene regulatory network we investigate is a chaotic motif task, as shown in Fig.~\ref{Fig2}{\bf b}. Chaotic motifs are minimal structures with simple interactions that can generate chaos in biological networks~\cite{zhang2012chaotic}. The chaotic dynamics are described by the following differential  equations,
\be
\begin{aligned}
      \frac{dp_1}{dt} &= -p_1+\frac{\kappa^{h_n}}{\kappa^{h_n}+p_3^{h_n}}\frac{p_2^{h_n}}{\kappa^h+p_2^h},\\
  \frac{dp_2}{dt} &= -p_2+\frac{\kappa^{h_n}}{\kappa^h+p_4^{h_n}},\\
  \frac{dp_3}{dt} &= -p_3+\frac{\kappa^{h_n}}{\kappa^{h_n}+p_2^{h_n}}\frac{p_3^{h_n}}{\kappa^{h_n}+p_3^{h_n}},\\
  \frac{dp_4}{dt} &= -p_4+\frac{p_1^{h_n}}{\kappa^{h_n}+p_1^{h_n}},
\end{aligned}
\label{eq:motif}
\ee
where $p_i$  ($p_i \in [0, 1]$) represents the expression level of $i$-th gene. The regulatory interactions of genes are modeled by Hill functions with the cooperativity exponent ${h_n}$ and the activation coefficient $\kappa$~\cite{tsai2008robust}. 
As in the last learning task, we discretize the dynamics with a timestep $\delta t= 0.035$, and use the configured quantum reservoir to predict $100$  forward steps.

The results are shown in Fig.~\ref{Fig2}{\bf b}. 
As the chaotic motif represents a more challenging task than the oscillatory network, our configured quantum reservoir computing only captures the dynamics of the first $80$ steps, with an NMSE of $10^{-4}$. However, for the final $20$ steps, the accuracy of the prediction decreases, having a sizable discrepancy (Methods).
It is worth noting that we use different parameters in Eq.~\eqref{eq:motif} to generate the training and test datasets. In the case of the chaotic behavior of the dynamics, the time sequences in the training and test datasets differ significantly due to chaos. Achieving accurate predictions for the first $80$ steps of chaotic motifs is a noticeable achievement, indicating that our configured quantum reservoir computing correctly models the chaotic features. Moreover, we confirm that the discrepancy can be further resolved by increasing just one more qubit (Supplementary Information).

%Though chaotic tasks are more difficult to predict, the reservoir's predictions on the tasks are almost consistent, with an NMSE of approximately $10^{-4}$ for the first 80 steps. 

%However, the error sharply increases in the last 20 steps. According to the prediction results in Fig.~\ref{Fig2} {\bf b}, we find two components that deviate from the target rapidly towards the end of the prediction. The chaotic sequence prediction is sensitive to errors. During the prediction, the current output becomes the next input, so the error accumulates. As a result, inaccuracies grow quickly with improper inputs.

With the successful applications of the configured quantum reservoir computing technique on inferring the dynamical trajectories of gene regulatory systems, it is expected that this method will serve as a generic approach for modeling biological systems. A single configured quantum reservoir has the capability to reproduce 
multiple complex processes, which may display oscillatory or chaotic features. This presents a new avenue for modeling the dynamics and revealing the underlying mechanism of intricate biological processes using quantum reservoirs.

\begin{figure*}[htp]
\includegraphics[width=.9\linewidth]{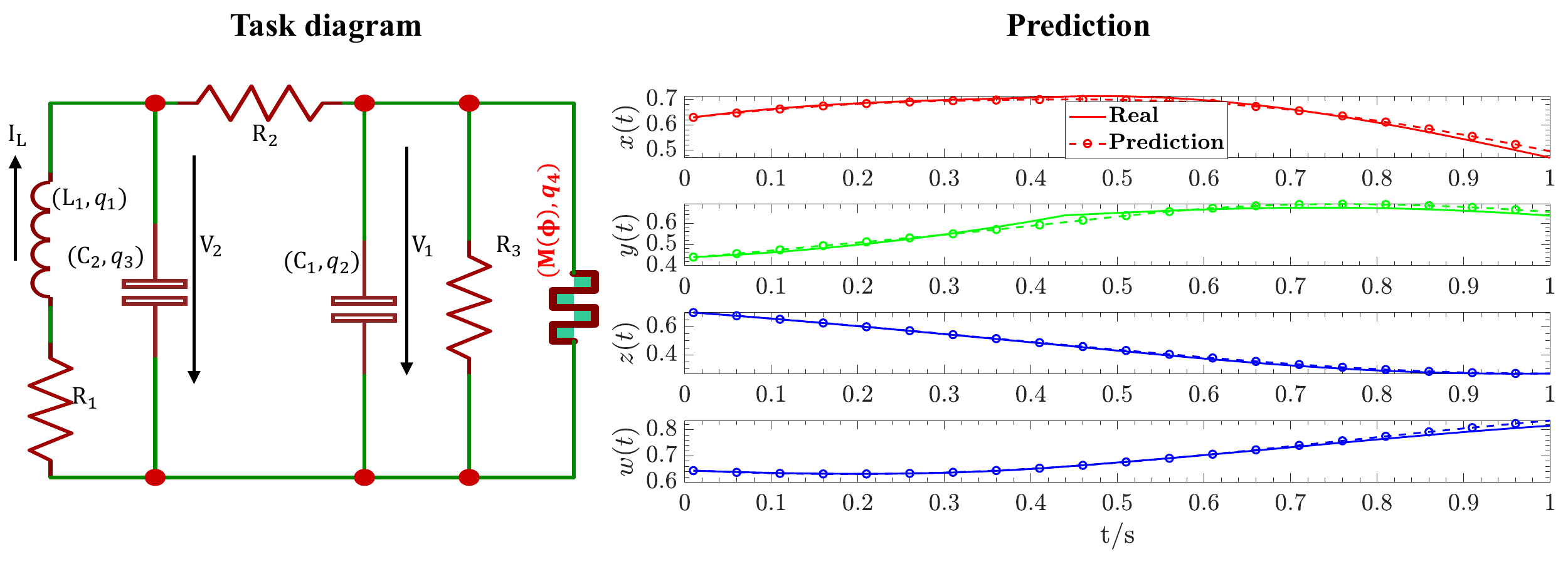}
\caption{{\bf Fractional-order Chua's circuit.} 
The figure shows the structure of the circuit and the prediction of the configured quantum reservoir computing.  The circuit dynamics are represented by four components, $x(t)$, $y(t)$, $z(t)$, and $w(t)$, whose time evolution are described by fractional differential equations. The dynamical variables $x$, and $y$  correspond to the voltage on the capacitors C$_1$ and C$_2$, in units of volts; $z$ and $w$ correspond to the current through the inductor L$_1$ in ampere, and $w$ the magnetic flux through the memristor in weber. 
} 
%We simulate $6100$ timesteps of the circuit dynamics. The first $100$ This particular task instance involves 6100 time steps, with the first 1000 steps excluded, the next 5000 steps used for training, and the final 100 steps reserved for reservoir prediction. For this task, we used a qubit number of $N=6$, a number of subintervals of $V=10$. }
\label{Fig3}
\end{figure*}

\subsection{Fractional-order Chua's circuit} 

We also apply the configured quantum reservoir computing to a fractional order Chua's circuit. 
The quantum reservoir applied here remains exactly the same as used for gene regulatory networks.
The memristor in this circuit (Fig.~\ref{Fig3}) provides nontrivial nonlinearity described by fractional derivatives (Methods), which could produce even more complex dynamics than conventional chaotic systems ~\cite{riewe1997mechanics,kiani2009chaotic,zhao2015novel}.  Fractional order chaos has generated much research interest~\cite{cafagna2008fractional,radwan2011stability,lu2005chaotic} for their applications in describing complex circuits ~\cite{petras2010fractional,Elwakil2010Fractional,Freeborn2013Survey} and fundamental distinction from the integer-order chaos ~\cite{Hartley1995Transactions}. 

The dynamics of the circuit in Fig.~\ref{Fig3} is characterized by  $x(t)$, $y(t)$, $z(t)$, and $w(t)$. Here,  $x$, and $y$ are the voltage on the capacitors C$_1$ and C$_2$, in units of volts, $z$ the current through the inductor L$_1$ in ampere, and $w$ the magnetic flux through the memristor in weber.
To construct the fractional order Chua's circuit, 
the electronic elements $C_1$, $C_2$, $L_1$, and the memristor, take fractional orders $q_1$, $q_2$, $q_3$, and $q_4$ (Methods). 
As a test example, we choose $R_1=100/130{\rm K} \Omega $, $R_2 = 100{\rm K} \Omega$, and $R_3=-200/3{\rm K} \Omega $, $L_1 = 10{\rm mH/s^{1-q_1}} $, $C_1=1{\rm \mu F/s^{1-q_2}}$, $C_2 = 10{\rm \mu F/s^{1-q_3}}$. 
The memristor $M(\phi)$ is a flux-controlled device, whose current ($I_M$) depends both its voltage ($V_M$) and flux ($\phi_M$). Its property is described by $I_M = f(\phi_M)V_M$ and $f(\phi)$ is a piecewise-linear function, $f(\phi) = 3{\rm \mu S\cdot s^{1-q_4}} ,\ \left |\phi\right|<1{\rm Wb}; f(\phi) = 8{\rm \mu S\cdot s^{1-q_4}},\ \left |\phi\right|>1{\rm Wb}$.
As shown in Fig.~\ref{Fig3}, the fractional Chua's circuit develops intricate nonlinear dynamics much more complex than standard LC circuits. This circuit has nontrivial features such as saturation of the  voltage on C$_2$, the non-monotonic dynamics of the voltage on C$_1$, and the anti-correlation between the current through L$_1$ and the magnetic flux through the memristor. 
Despite the complexity,  our configured quantum reservoir computing is capable of producing quantitatively correct dynamics upto one second.  The nontrivial features of the circuit have been correctly captured by the quantum reservoir. 
The NMSE for this task of learning fractional order chaos still reaches $10^{-4}$. 
This further demonstrates the exceptional learning capacity of our configured quantum reservoir computing.

\begin{figure*}[htp]
\includegraphics[width=.85\linewidth]{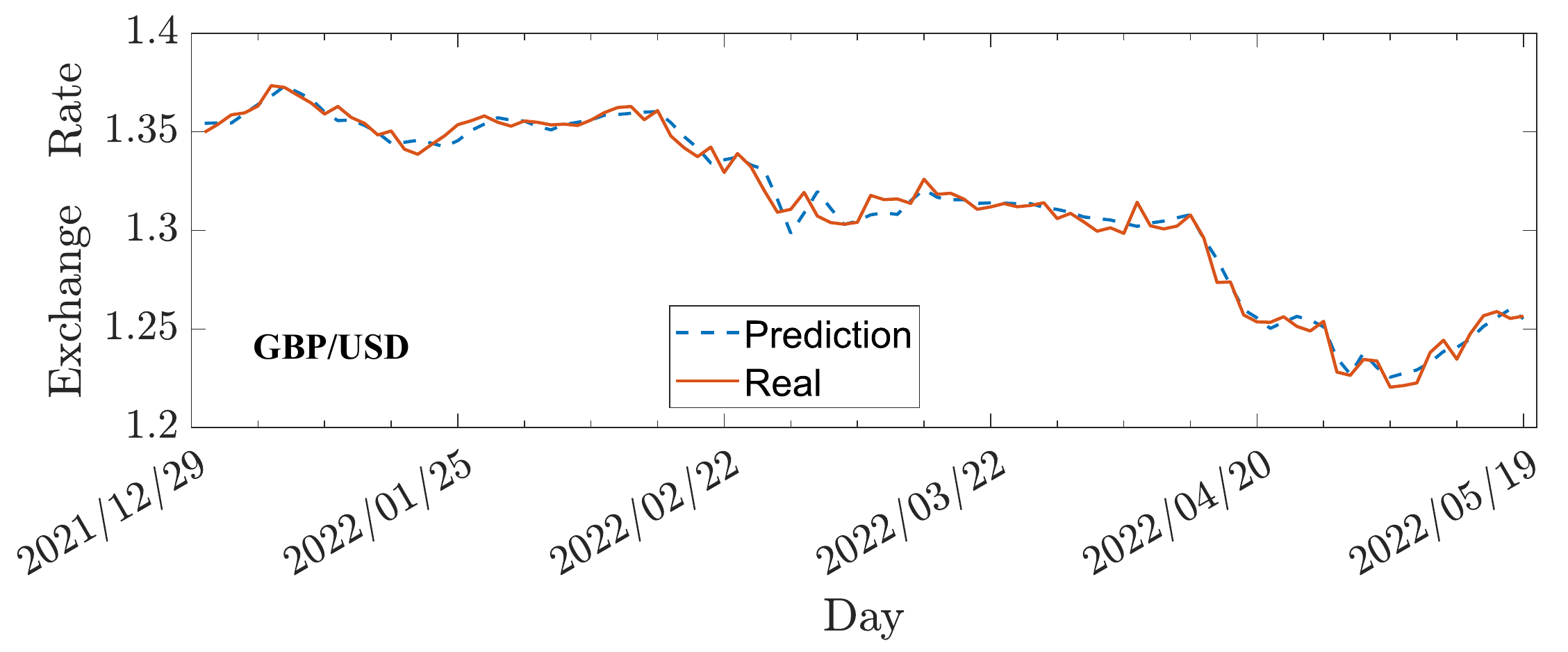}
\caption{ {\bf FX market forecast}. Here, we present the prediction results for GBP/USD by configured quantum reservoir computing.
The quantum reservoir is trained on the AUD/USD and  NZD/USD exchange rates  in the period from February 8, 2018 to May 19, 2022, and is then applied to GBP/USD. The solid and dotted lines represent the actual exchange rate and the quantum reservoir prediction, respectively. In this application, we use $8$ qubits for the  quantum reservoir.
The relative error of our prediction is $\sqrt{{\rm NMSE}} \approx 0.3\%$. 
This is already significantly smaller than the day-to-day fluctuations, which are about $2\%$. 
}
\label{Fig4}
\end{figure*}

\subsection{FX market forecast}   
To demonstrate the configured quantum reservoir computing also applies to stochastic time sequence predictions, we investigate the applications on the FX markets. For the complexity induced by stochastic fluctuations that  make this task drastically different from those deterministic tasks studied above, we choose the qubit number $n=8$, and retrain the quantum reservoir using the FX market data. 
We adopt a sliding window approach that has been developed in using reservoir computing for Fintech tasks~\cite{wang2021stock}.  The price at day $t+1$ to predict is modeled as an output ${\bf y}_k$ with a dimension $d_{\rm out} = 1$. The prices in the $6$ prior trading days are taken to form  an input signal, ${\bf s}_k$, with a dimension $d_{\rm in}  = 6$. The reservoir parameters ($\theta$) are trained according to the AUD/USD and  NZD/USD exchange rates in the period from February 8, 2018 to May 19, 2022 (Eq.~\eqref{eq:costfunction}). The learning performance is tested on GBP/USD (Fig.~\ref{Fig4}) from February 12, 2022 to May 19, 2022. The configured quantum reservoir prediction has reasonable accuracy, and it closely reproduces the movement of the ground truth curve. The corresponding NMSE reaches {\blue $10^{-5}$}, which means the relative error of our prediction is $\sqrt{{\rm NMSE}} \approx 0.3\%$. 
Despite having  only eight qubits in the quantum reservoir, our prediction exhibits one-order-of-magnitude improvement in accuracy when compared to previous studies using  classical reservoir computing with even more than one hundred reservoir nodes~\cite{wang2022echo}.
% This is comparable to classical reservoir computing applications to GBP/USD forecasting using $100$ reservoir nodes, although only $6$ qubits are used here. 
It is worth remarking here that the  day-to-day fluctuations  of the GBP/USD exchange rate are about {$2\%$}. This implies that quantum reservoir computing potentially  creates room for significant arbitrage if  more quantum computing resources are provided. 

We also carry out alternative tests where we take two of AUD/USD, NZD/USD, and GBP/USD exchange rates as training data, and the other one for testing. The resultant prediction accuracy is at the same level as presented in Fig.~\ref{Fig4} (Supplementary Information).

\begin{figure*}[htp]
\includegraphics[width=.9\linewidth]{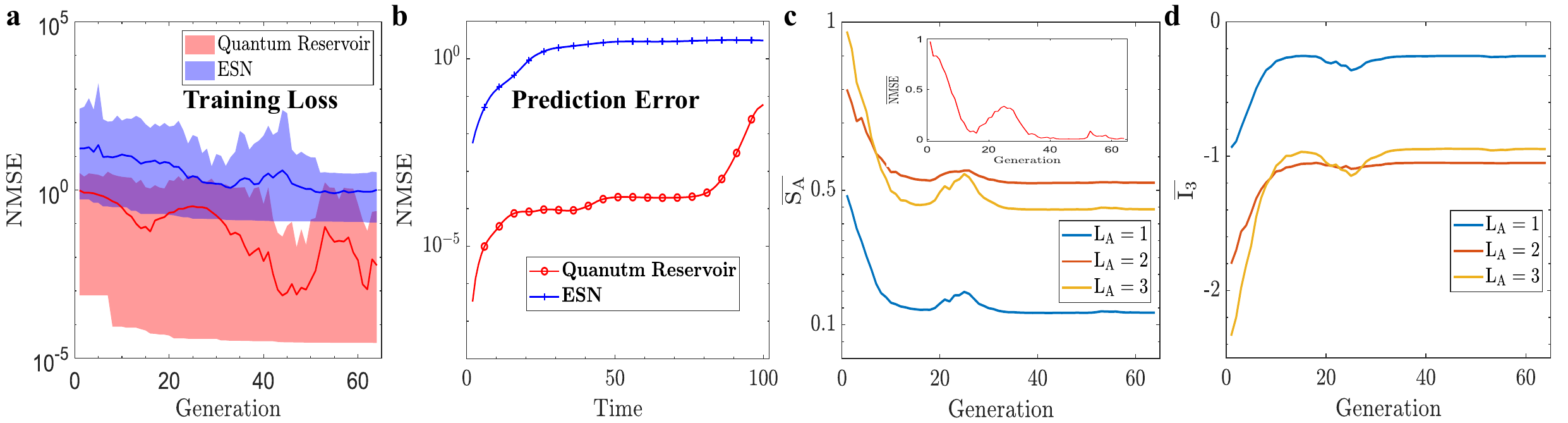}
\caption{ {\bf The emergent quantum advantage with configured quantum reservoir computing}. 
{\bf a}, the training loss during the optimization iteration by the genetic algorithm. The solid lines represent the averaged value, and the shaded region surrounding them illustrates the distribution of different reservoir configurations searched by the genetic algorithm. 
{\bf b}, the reservoir prediction accuracy on the testing dataset. 
{\bf c}, the bipartite entanglement entropy of the eigenstates of the quantum reservoir Hamiltonian. {\bf d}, the tripartite mutual information ${\rm I_3(A,C,D)}$ of the unitary time evolution operator of the quantum reservoir. In {\bf c} and {\bf d}, we average over the searched reservoir configurations. Here, we choose $N_{\rm node} = 6$ for ESN, and qubit number $n=6$ for the quantum model. Both classical and quantum reservoirs are trained and applied to gene regulatory networks and fractional order Chua's circuit models.
The setting of the genetic algorithm for optimization is identical for them, for the fairest comparison.}
\label{Fig5}
\end{figure*}

\section{The emergent quantum advantage}
We have demonstrated outstanding performance in prediction accuracy and transferability through the implementation of the configured quantum reservoir computing in the aforementioned learning tasks. In order to characterize the quantum effects in the learning process, we conduct  a direct comparison with ESN, a prevalent classical reservoir computing method, where we observe considerable quantum advantage. We attribute the exceptional learning capability of the configured quantum reservoir computing to quantum coherence, which is validated by constructing synthetic models that allow for control of the degree of quantum coherence.

\begin{figure*}[htp]
\includegraphics[width=.9\linewidth]{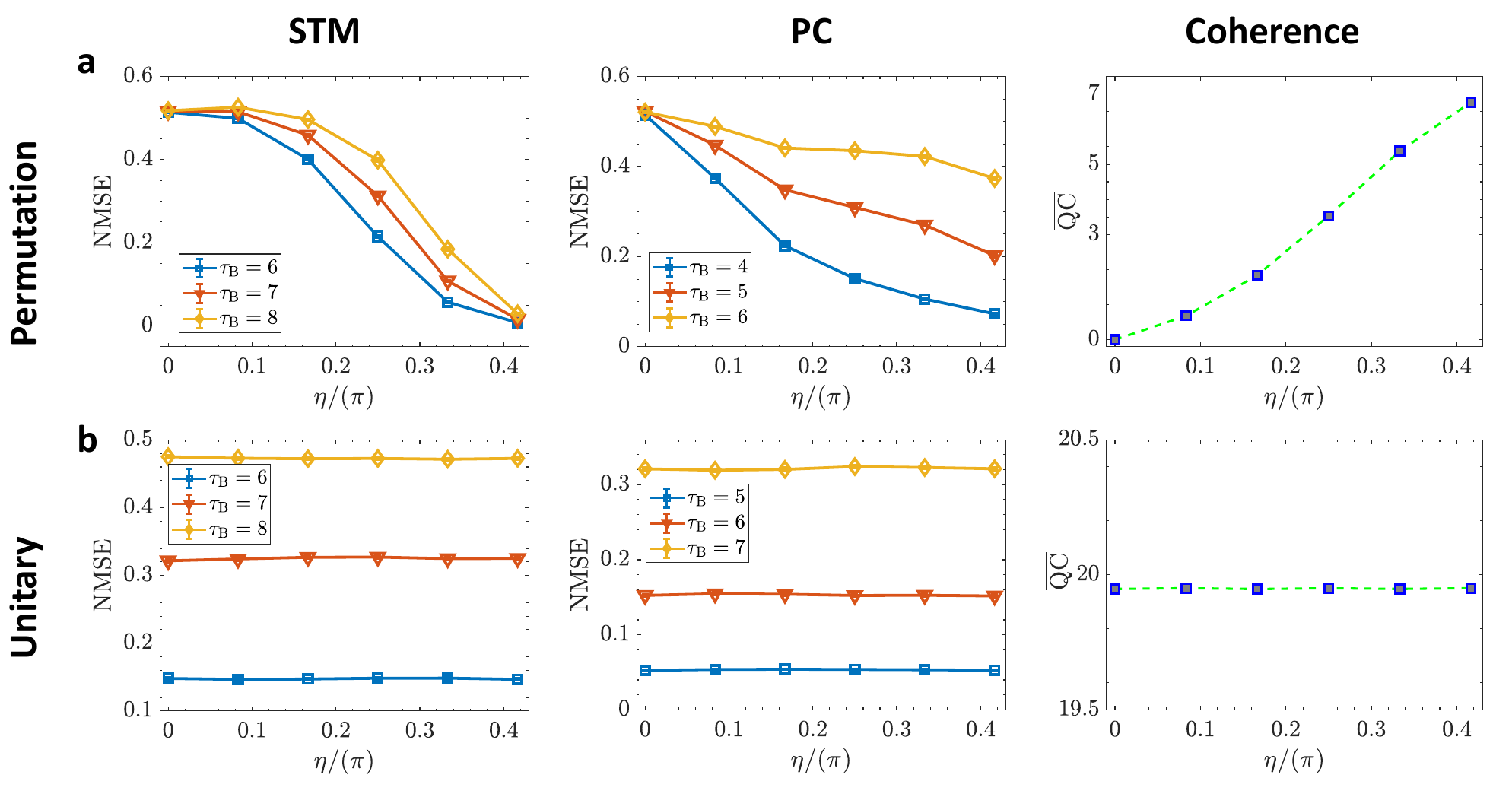}
\caption{ {\bf The origin of quantum advantage}. 
We study the computing performance of two types of reservoirs, random permutation and random unitary (see the {\it main text}).
For the former, the degree of quantum coherence (QC) systematically increases as we increase the encoding angle ($\eta$) from $0$ to $\pi/2$. For the latter, it stays constant. {Here, QC is averaged over the searched reservoir configurations and different time steps}.  Reservoir computing is performed on short-term memory (STM) and parity check (PC) tasks.
We choose the qubit number $n=8$ here. 
We observe systematic improvement in the prediction accuracy with increasing $\eta$ for the random permutation model, whereas the performance remains roughly the same for the random unitary model. 
}
\label{Fig6}
\end{figure*}

\subsection{Comparison with classical reservoir computing}  
ESN is a widely used  classical reservoir computing model. Having $N_{\rm node}$ number of reservoir nodes, its nonlinear dynamics is described by the evolution of a $N_{\rm node}$-dimensional vector ${\bf x}_k$, 
\be
      {\bf x}_k = \tanh ({\bf M} \cdot {\bf x}_{k-1}+ {\bf D} \cdot {\bf s}_k). 
\label{eq:reservoirdynamics}
\ee 
The output is defined by, 
$
     {\bf y}_k = {\bf W} \cdot {\bf x}_k + {\bf B}. 
$
In defining the nonlinear reservoir dynamics in Eq.~\eqref{eq:reservoirdynamics},  
${\bf D}$ is an encoding matrix ($N_{\rm node} \times d_{\rm in}$) with matrix elements  randomly chosen between $-1$ and $1$~\cite{Keisuke2017,jaeger2004harnessing}, 
and  the coupling matrix ${\bf M}$ contains $N_{\rm node}^2 $  reservoir parameters. These reservoir parameters are configured with the same genetic algorithm using the same setting as the configured quantum reservoir computing for a fair comparison.  
The only difference in ESN from quantum reservoir computing is that the reservoir dynamics are classical.

In Fig.~\ref{Fig5}, we choose $N_{\rm node} = 6$ (equal to the number of qubits) for ESN. 
The number of reservoir parameters in the classical model is about two times of the quantum model for the learning tasks.
ESN is performed on the deterministic learning tasks described in Sec.~\ref{sec:Qapp}. 
Fig.~\ref{Fig5}{\bf a} shows the training loss, namely the ${\rm NMSE}$ on the training dataset. In the training iteration by the genetic algorithm, although the decrease in the training loss of ESN is somewhat more systematic than the quantum model, the optimal training loss (${\cal L}_{\rm opt}$) for the latter is considerably lower. For the classical model, we have ${\cal L}_{\rm opt} = 0.1$, and for the quantum model, ${\cal L}_{\rm opt} = 2.8\times 10^{-5}$. 
The comparison in the prediction accuracy on the testing dataset is also dramatic (see Fig.~\ref{Fig5}{\bf b}). For an extended period of evolution time, the prediction of the quantum model is four orders of magnitude more accurate than the classical model. We also examine the performance of ESN with a much larger number of reservoir nodes, upto $N_{\rm node} = 120$ (see Supplementary Information). The training loss can be improved to $5.0\times 10^{-6}$, but the prediction accuracy is still much (four orders of magnitude) worse than our configured quantum reservoir computing. This implies that learning with the quantum model is significantly more transferable than the classical approach. These results demonstrate an affirmative advantage in the configured quantum reservoir computing over the corresponding classical approach in multi-task machine learning. The quantum approach has a surprisingly larger degree of transferability.

To characterize the quantum correlation effects present in the reservoir Hamiltonians, we provide the entanglement entropy $S_A$~\cite{lydzba2020eigenstate} of the Hamiltonian eigenstates and the tripartite mutual information $I_3(A; C,D)$~\cite{seshadri2018tripartite,shen2020information} of the unitary time evolution operator $\exp{(-i\hat{H}(\theta)\tau)}$ in Fig.~\ref{Fig5}({\bf c}, {\bf d}). In the calculation, we choose the subregion $A$ to be the first one, two, or three qubits, the subregion $C$ the same as $A$, and $D$ the rest of the reservoir system. 
The training of the quantum reservoir starts from randomly initialized non-local Hamiltonians (Eq.~\eqref{eq:Hamiltonian}). These models have efficient information scrambling power but have limited memory storage as the reservoir would quickly undergo thermalization~\cite{xia2022reservoir}. During the training process, a rapid improvement in the training loss is observed at the early stages, accompanied by an almost linear increase in the tripartite information and a linear decrease in the average eigenstate entanglement. This suggests that the quantum reservoir becomes less scrambled and non-thermal. The quantum correlations as measured by entanglement entropy and mutual information saturate at the late stage of the training. 
Prior to that, the training loss develops a notable bump, which correlates with a rise in the entanglement entropy and a drop in the mutual information. This confirms the learning power of the quantum reservoir is  indeed closely related to  the intricate quantum correlation effects embedded within the quantum reservoir.

\subsection{The origin of quantum advantage} 

In previous studies on quantum speedup, it has been established that the quantum advantage exhibited by various quantum algorithms stems from quantum coherence.
It has been shown to be an essential resource for the Deutsch-Jozsa algorithm~\cite{Hillery2016Coherence}, as well as a crucial factor in quantum amplitude amplification, which leads to the quadratic speedup in Grover search~\cite{Shi2017Coherence,anand2016coherence}. 
In order to understand the superior learning power of our configured quantum reservoir computing, we seek to investigate the quantitative impact of quantum coherence on learning performance.

We construct a quantum reservoir computing model in which we can systematically adjust the degree of quantum coherence. 
Here we consider learning tasks with $d_{\rm in} = d_{\rm out} = 1$ for simplicity. 
An encoding angle $\eta$ is introduced to control the degree of quantum coherence. 
The one-dimensional input signal $s_k$ is injected to the reservoir by measuring the first qubit followed by resetting it to $\ket{\psi_{s_k}} = \sqrt{s_k}\ket{\eta}_{+}+\sqrt{1-s_k}\ket{\eta}_{-}$, with 
$\ket{\eta}_{\pm}$ the two eigenstates of $\sin \eta \hat{\sigma}_x + \cos\eta \hat{\sigma}_z$. 
The reservoir dynamics is generated by random permutation of the computation basis, implemented by performing ten exchanges of basis states between two successive input signal injections. 
With the encoding angle $\eta = 0$, this process does not produce any quantum entanglement and consequently, 
the reservoir state remains separable. 
In contrast, for $0<\eta<\pi/2$,
the reservoir state contains coherent quantum superposition among the computation basis states, and the permutation process creates sufficient quantum entanglement, 
{making the reservoir no longer separable.}  
Quantitatively, the degree of quantum coherence is measured by the $l_1$ norm  of the off-diagonal part of the density matrix $\rho_{\rm od}$ as, 
\be 
{\rm QC} \equiv \norm{\rho_{\rm od}}_{l_1}.
\label{eq:QC}
\ee

The performance of the random permutation model is examined on short-term memory and parity check, two standard reservoir computing tasks widely used for benchmarking the learning capacity~\cite{bertschinger2004rea,nakajima2014exploiting}. 
The corresponding time sequence functions are $y_k^{STM} = s_{k-\tau_B}$ and $y_k^{PC} = (\sum_{q=0}^{\tau_B}s_{k-q})\text{ mod }2$, where $\tau_B$ is the time delay and $s_k$  takes binary values of $0$ or $1$. 
We average over $100$ random instances for $s_k$, and simulate $5000$ time steps. The first $1000$ steps are used for the washout~\cite{Keisuke2017}, and the next $3000$ steps are taken for determining the weights of the linear model (Sec.~\ref{sec:theory}). The final $1000$ steps are reserved for testing the reservoir performance. 
As shown in Fig.~\ref{Fig6}{\bf a}, the learning performance monotonically increases for both short term memory  and parity check, as we increase the encoding angle $\eta$. 
This improvement is systematic for various choices of time delay. The systematic improvement of the learning performance correlates with the degree of quantum coherence in the quantum reservoir.

As a comparative study, we also examine the performance of a random unitary quantum reservoir on the same learning tasks.
The random unitary model is deliberately set to be the same as the random permutation, except the procedure of basis state exchange is replaced by a Haar random unitary. The quantum reservoir then involves sufficient quantum superposition, irrespective of the encoding angle. With the random unitary model, we find that the learning performance on short term memory and parity check remains more or less unaffected by increasing the encoding angle (Fig.~\ref{Fig6}({\bf b})), which is consistent with the fact of quantum coherence being constant. 

Based on our findings with the random permutation and random unitary models, we attribute the superior learning capability of quantum reservoir computing to the presence of quantum coherence in the quantum reservoir system. This suggests that the complexity of quantum many-body systems, which cannot be simulated efficiently by classical computing, offers valuable resources for machine learning applications.

\section{Conclusion}
We have presented a novel approach to quantum reservoir computing, which outperforms classical reservoir computing in multi-task machine learning. Our approach has been demonstrated on gene regulatory networks and fractional order Chua's circuits, where the quantum approach with six qubits achieved comparable performance to classical reservoirs with hundreds of nodes on the training dataset, but four orders of magnitude higher accuracy on the testing dataset. Furthermore, our approach shows significant improvement in the prediction accuracy of FX market forecasts compared to previous reservoir computing studies. These results highlight the potential of configured quantum reservoir computing to achieve quantum advantage in NISQ devices, which exhibit complex quantum dynamics that are not efficiently simulatable by classical resources. We attribute the superior computation power of our approach to the quantum coherence embedded in the quantum reservoir dynamics. Overall, our findings offer a promising avenue for quantum-enhanced machine learning with practical applications.

%This work proposes the use of a genetic algorithm to systematically improve the learning performance of quantum reservoirs. We demonstrate the effectiveness of the configured quantum reservoir on various tasks, including a synthetic oscillatory network of transcriptional regulators, chaotic motifs in gene regulatory networks, and a fractional-order Chua's circuit with a memristor. The quantum reservoir successfully learns all three tasks simultaneously and achieves accurate predictions. We also apply the configured quantum reservoir to FX market applications, and it yields reasonable prediction accuracy.

%Furthermore, we investigate the underlying mechanism of the configured quantum reservoirs. We find that during training, the reservoir becomes less entangled and scrambling to balance its memory and nonlinearity. Comparing our configured quantum reservoirs to ESNs, we demonstrate the quantum advantage of our approach. To study the origin of this quantum advantage, we design two additional reservoirs: a random unitary and a random permutation, where the former generates quantum coherence, but the latter is not able to generate quantum coherence. We further introduce quantum coherence quantitatively by rotating the encoding bases and observe that higher quantum coherence leads to better performance of the quantum reservoir and conclude that quantum coherence is the origin of the configured quantum reservoir's quantum advantage.

\section{Methods}

\subsection{Encoding protocol} 
The quantum reservoir dynamics is described by a density matrix $\rho (t)$ in the computation basis (the Pauli-$\hat{\sigma}^Z$ eigenbasis). At the time $t=0$, the quantum dynamics starts from an infinite temperature state with $\rho(0) = \mathbbm{1}/2^{n}$. At each sequential injection of the input signal as labeled by $k$, the first $d_{\rm in}/2$ qubits are measured in the computation basis, and then reset to   
\be 
\otimes_{j=1}^{d_{\rm in}/2} \left[  \sqrt{1-s_k(2j-1)}\ket{+}+e^{-is_k(2j)}\sqrt{s_k(2j-1)}\ket{-} \right],
\ee 
with $j$ indexing the qubits, $s_k(\ldots)$  the elements of the ${\bf s}_k$ vector, 
and $\ket{\pm}$ the eigenstates of the Pauli-$\hat{\sigma}^X$ operator. 
The performance of using a different basis for encoding (Pauli-$\hat{\sigma}^Z$ basis) is provided in Supplementary Information. 
The information of the input signal is thus encoded in the amplitude and the phase degrees of freedom of the reservoir qubits upon the injection. 
During each time interval ($\tau$) between the successive injections, the quantum reservoir is let evolve according to a parameterized Hamiltonian [$\hat{H}(\theta)$], 
\be 
\hat{\rho}(k\tau+\Delta t) = e^{-i\hat{H}(\theta) \Delta t} \hat{\rho} (k\tau) e^{i\hat{H}(\theta)\Delta t}.
\ee

For the readout, each time interval, $\tau$, is further split into multiple ($V$) subintervals, and the measurements are performed in the Pauli-$\hat{\sigma}^X$, $\hat{\sigma}^Y$, and $\hat{\sigma}^Z$ basis at the end of each subinterval, to build the final quantum reservoir output. 
The measured quantum expectation values correspond to a result tensor 
${\cal A}_{k, v,j,a} =({\rm Tr}[\hat{\rho}(k\tau+\tau v/V)\hat{\sigma}_j^a])$, with $v$, $j$, and $a$ indexing the subintervals, qubits, and the three Pauli operators, respectively. For computation convenience, the result tensor is flattened into a vector, ${\bf A}_k$. The  final output is produced by acting a linear regression model on ${\bf A}_k$.   

\subsection{Determination of the weights of the linear regression model}
In both the quantum reservoir training and the applications to the learning tasks, the weights of the linear regression model are determined by minimizing 
$\sum_k \left( {{\bf y}_k-{\bf y}_k^\star}\right)^2$. 
More specifically, the time sequences are divided into three groups, $k<G_0$, and $G_0\leq k<G_1$, and $G_1\le k<K$. 
The time sequences in the first group are used to guide the quantum reservoir dynamics following the standard washout technique in reservoir computing~\cite{Keisuke2017,jaeger2004harnessing}.  
The data in the second group is used to determine the weights of the linear regression model, which is efficiently performed by matrix multiplication~\cite{Keisuke2017}. 
The third group is used to characterize the performance of the quantum reservoir by computing the NMSE. In training the quantum reservoir, the NMSE for the third group is taken as the cost function to configure the reservoir parameters $\theta$. 

For the learning tasks of gene regulatory networks and fractional order Chua's circuit, we use $G_0 = 1000$, and $G_1 = 6000$, $K = 6100$.
For the FX market forecast, we use $G_0 = 200$, $G_1 = 1000$, and $K= 1100$.

\subsection{Training of the quantum reservoir by a genetic algorithm} 
With the cost function determined as described above, the quantum reservoir parameters $\theta$ are configured accordingly by a standard genetic algorithm (GA). 
The GA is initialized with a population size of $200$. Half of the population is generated  randomly by sampling the Hamiltonian parameters in Eq.~\eqref{eq:Hamiltonian},  and the other half is obtained from Refs.~\onlinecite{Keisuke2017, martinez2021dynamical, xia2022reservoir, kutvonen2020optimizing}, with $25$ initial populations provided for each type of  four quantum reservoir models from the previous literatures. 
Alternatively, we can also randomly initialize the population without invoking any prior knowledge, which we also have examined and found no significant difference (Supplementary Information).

\subsection{Training and testing datasets}
For all the deterministic learning tasks, the training and testing datasets are obtained by solving the differential equations using  the 4-th order Runge Kutta method. The numerical results are normalized to fit in the window of $[0,1]$, in order to treat different tasks on equal footing.

For the oscillatory gene regulatory network as described by Eq.~\eqref{eq:Oscillatory}, the training data is generated by taking $\alpha = 400$, $\alpha_0 = 0.4$, $h = 2$, and $\beta = 5$. 
The testing data is generated using a different set of parameters, $\alpha = 500$, $\alpha_0 = 0.5$, $h = 2$, and $\beta = 5$. We choose a time step $\delta t = 0.05$ in generating the time sequences. 
For the chaotic motif gene regulatory network (Eq.~\eqref{eq:motif}), the parameters for the training data are $h = 2.5$ and $k = 0.134$, and the testing data are $h = 2.49$ and $k = 0.135$. The time step used for this learning  task is $\delta t = 0.035$.  
For both the oscillatory and the chaotic motif gene regulatory networks, we generate a total number of $6100$ time steps. 

The fractional order Chua's circuit is described by 
\be
\begin{aligned}
{}_0D^{q1}_t x(t) &= \alpha(y(t)-x(t)+\zeta x(t)- f(\phi_M)x(t)), \\
{}_0D^{q2}_t y(t) &= x(t)-y(t)+z(t),\\
{}_0D^{q1}_t z(t) &= -\beta y(t) - \gamma z(t),\\
{}_0D^{q1}_t w(t) &= x(t),
\end{aligned}
\ee
where $q_1$, $q_2$, $q_3$, and $q_4$ are the fractional orders determined by the circuit. 
%$x$ and $y$ are the voltage of capacitance $C_1$ and $C_2$, respectively. $z$ is the current of inductance $L_1$, and $\phi$ is the flux of memristor $M$. 
For both training and testing datasets, we choose $q_1=q_2=q_3=q_4=0.97$~\cite{petras2010fractional}. 
The rest of the parameters are given by $\alpha = 1/C_2$, $\beta=1/L_1$, $\gamma= R_1/L_1$, and $\zeta = 1/R_3$. For the training dataset, we choose $R_1=100/130{\rm K} \Omega $, $R_2 = 100{\rm K} \Omega$, and $R_3=-200/3{\rm K} \Omega $, $L_1 = 10{\rm mH/s^{1-q_1}} $, $C_1=1{\rm \mu F/s^{1-q_2}}$, $C_2 = 10{\rm \mu F/s^{1-q_3}}$,
and $f(\phi) = 3{\rm \mu S\cdot s^{1-q_4}} ,\ \left |\phi\right|<1{\rm Wb}; f(\phi) = 8{\rm \mu S\cdot s^{1-q_4}},\ \left |\phi\right|>1{\rm Wb}$. For the testing dataset, $f(\phi) = 8{\rm \mu S\cdot s^{1-q_4}},\ \left |\phi\right|>1{\rm Wb}$ is replaced by $f(\phi) = 7{\rm \mu S\cdot s^{1-q_4}},\ \left |\phi\right|>1{\rm Wb}$.
%$\alpha = 10 $, $\beta = 13$, $\gamma = 0.1 $, $\zeta = 1.5$, $a= 0.3$, and $b = 0.8$, while in the testing data, the parameter $b$ has been changed to $b = 0.7$. 
In numerical simulations, we implement the method of solving fractal order differential equations in Ref.~\onlinecite{petras2010fractional}. 
For this task, we choose  $\delta t = 0.01$s, and also generate $6100$ time steps. 

For all the three learning tasks described by differential equations, their solutions at time step $k$ define the input sequence ${\bf s}_k$, and the solutions at the next step, i.e., $k+1$ define the output sequence ${\bf y}_k ^\star$.  
This setting is designed for the reservoir to predict forward evolution of time sequences. 
In testing the reservoir performance, the input signal ${\bf s}_k$ at $k>G_1$ is set to be the reservoir predicted output ${\bf y}_{k-1}$.

For the FX market forecast, the training dataset contains the exchange rates of  USD/CHF, NZD/USD, and AUD/USD in the period from February 8, 2018 to May 19, 2022.  
The testing dataset contains GBP/USD from February 12, 2022 to May 19, 2022.   
The exchange rates are normalized to treat them on equal footing. The normalized data is then denoised by a discrete wavelet transform technique widely used in 
stock market forecast~\cite{liu2022forecasting}.  
In testing the performance of our quantum reservoir computing, we use the raw data without normalization or denoising in calculating the NMSE for the prediction in Fig.~\ref{Fig4}.

\section{Acknowledgement}

We acknowledge helpful discussion with Andrew Chi-Chih Yao, Xun Gao, and Huangjun Zhu. 
This work is supported by National Program on Key Basic Research Project of China (Grant
No. 2021YFA1400900), National Natural Science Foundation of China (Grants No. 11934002, 12075128, T2225008), Shanghai Municipal Science and Technology Major Project (Grant No. 2019SHZDZX01), and Shanghai Science Foundation
(Grants No.21QA1400500).

\bibliography{references}

\clearpage

\begin{widetext}
\renewcommand{\theequation}{S\arabic{equation}}
\renewcommand{\thesection}{S-\arabic{section}}
\renewcommand{\thefigure}{S\arabic{figure}}
\renewcommand{\thetable}{S\arabic{table}}
\setcounter{equation}{0}
\setcounter{figure}{0}
\setcounter{table}{0}
\setcounter{section}{0}
% \newpage

\begin{center}
{\Huge \bf Supplementary Information} \\
\end{center}

\section{FX market forecast}
In the main text, we have presented the prediction results of GBP/AUD. Here, we carry out alternative tests where we take two
of AUD/USD, NZD/USD, and GBP/USD exchange rates as training data, and the other one for testing. The corresponding results are illustrated in Fig.~\ref{FigS1}. The exchange rates are in the period from February 8, 2018 to May 19, 2022 and the learning performance is tested from February 12, 2022 to May 19, 2022.

Our configured quantum reservoir prediction exhibits reasonable accuracy for both NZD/USD and AUD/USD, closely mimicking the actual curve. In both cases, the corresponding NMSE value is $10^{-5}$, indicating that our prediction has a relative error of approximately 0.3$\%$ (i.e., $\sqrt{{\rm NMSE}} \approx 0.3\%$). It is  noteworthy that the three exchange rates experience daily fluctuations of approximately $2\%$. This suggests that with more quantum computing resources, quantum reservoir computing could potentially  provide a significant opportunity for arbitrage.

\begin{figure}[htp]
\includegraphics[width=1\linewidth]{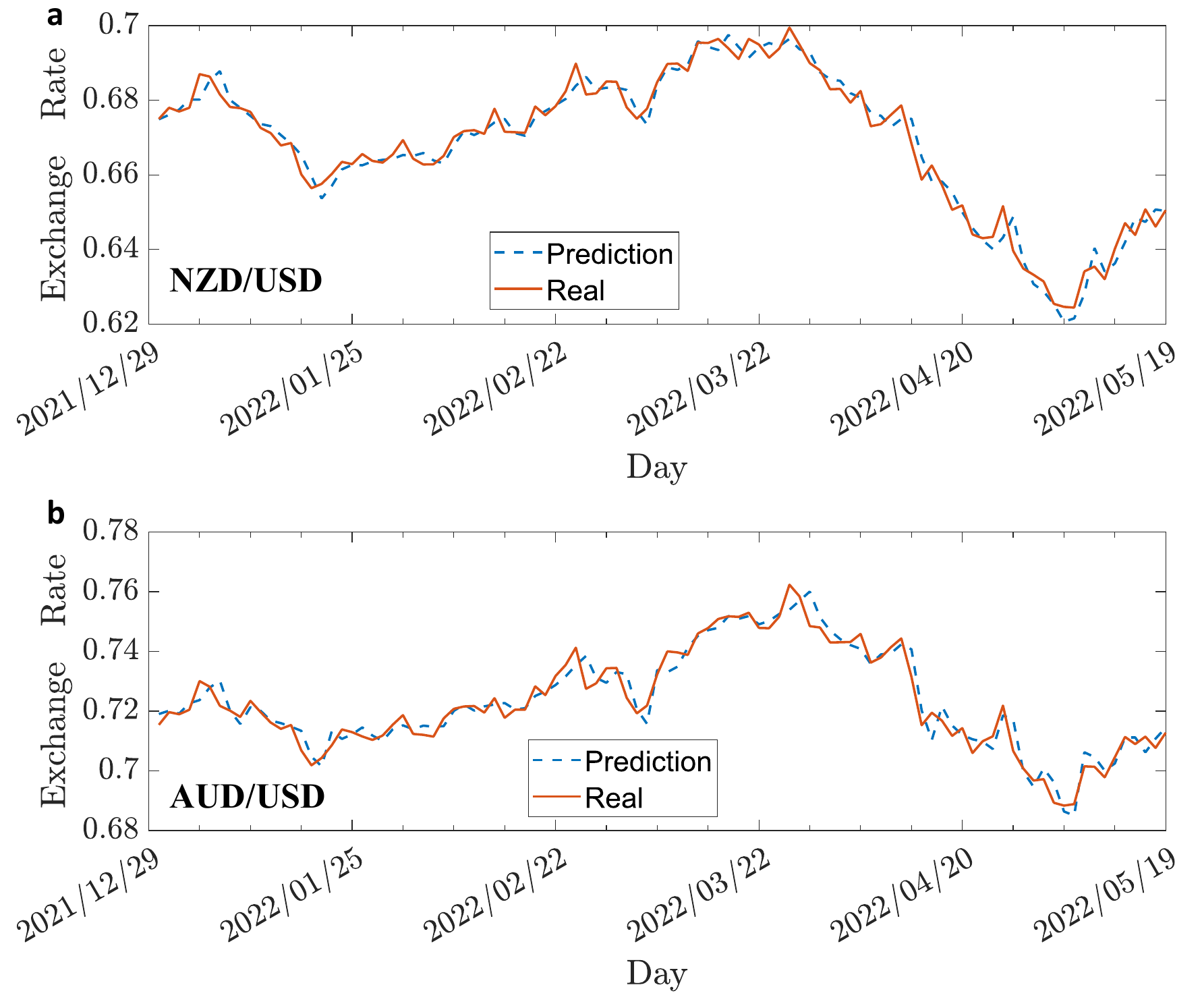}
\caption{{\bf FX market forecast}. Here, we present the prediction results for NZD/USD and AUD/UD by configured quantum reservoir computing. Here, we take two of AUD/USD, NZD/USD, and GBP/USD exchange rates as training data, and the other one for testing. 
The solid and dotted lines represent the actual exchange rate and the quantum reservoir prediction, respectively. In this application, we use $8$ qubits for the  quantum reservoir.}
\label{FigS1}
\end{figure}

\section{The performance of classical reservoir computing}
In the main text, we set the number of nodes in the Echo State Network (ESN) to be $N_{\rm node}=6$, which is equal to the number of qubits. 
Our configured quantum reservoir computing is four orders of magnitude more accurate than the classical model for both training and prediction errors. In this section, we investigate the effect of increasing $N_{\rm node}$ and vary the spectral radius ${r}$ of the coupling matrix {\bf M}, where the spectral radius ${r}$ is the maximal eigenvalue of {\bf M}. It has been previously reported that the computational power of ESN is closely related to the spectral radius of the coupling matrix~\cite{jaeger2004harnessing,verstraeten2007experimental}. We configure the coupling matrix {\bf M} using the same genetic algorithm and settings as the configured quantum reservoir computing. In our study, we set ${r}$ to $\{0.7, 0.8, 0.9\}$ and generate the initial population of the genetic algorithm based on ${r}$, but we do not constrain the spectral radius during training. We perform ESN on the deterministic learning tasks described in the main text, and Fig.~\ref{FigS2}{\bf a} shows the training loss, i.e., the ${\rm NMSE}$ on the training dataset. In the training iteration by the genetic algorithm, the decrease
in the average training loss of ESN is no longer systematic but the optimal training loss (${\cal L}_{\rm opt}$) is considerably lower. For the ${N_{\rm node} = 60}$ case, ${\cal L}_{\rm opt}$ is $6.99\times 10^{-6}$, where ${\rm r}$ takes the value $0.9$. For the ${N_{\rm node} = 120}$ case, the ${\cal L}_{\rm opt}$ is $5.01\times 10^{-6}$, where ${\rm r}$ takes the value $0.8$. In the main text, we reported the optimal training loss ${\cal L}_{\rm opt} = 0.1$ with ${N_{\rm node} = 6}$. Hence, we observe that the optimal training loss decreases with the increase of ${ N_{\rm node}}$.

Although the training loss of the ESN with $N_{\rm node} = 120$ becomes smaller than the configured quantum reservoir computing model (qubit number $n =6$), 
%In the classical model, ${\rm N_{node}}$ is set to 20 times that of the configured quantum reservoir computing model, and the optimal training loss is only one order of magnitude smaller than that of the configured quantum reservoir computing with $n = 6$. Despite having a lower optimal training loss, 
the classical model with a large ${N_{\rm node}}$ still has much worse prediction accuracy (four orders of magnitude) than our configured quantum reservoir computing model for a large time, as shown in Fig.~\ref{FigS2}{\bf b}. Increasing ${N_{\rm node}}$ from 60 to 120 does not improve the prediction accuracy much, and even leads  to worse performance for longer times, indicating that the classical model lacks transferability even with a large $N_{\rm node}$. In contrast, quantum reservoir computing exhibits a surprisingly high degree of transferability even with a small number of qubits $n$. 
%This is due to quantum coherence, as demonstrated in the main text.

\begin{figure}[htp]
\includegraphics[width=1\linewidth]{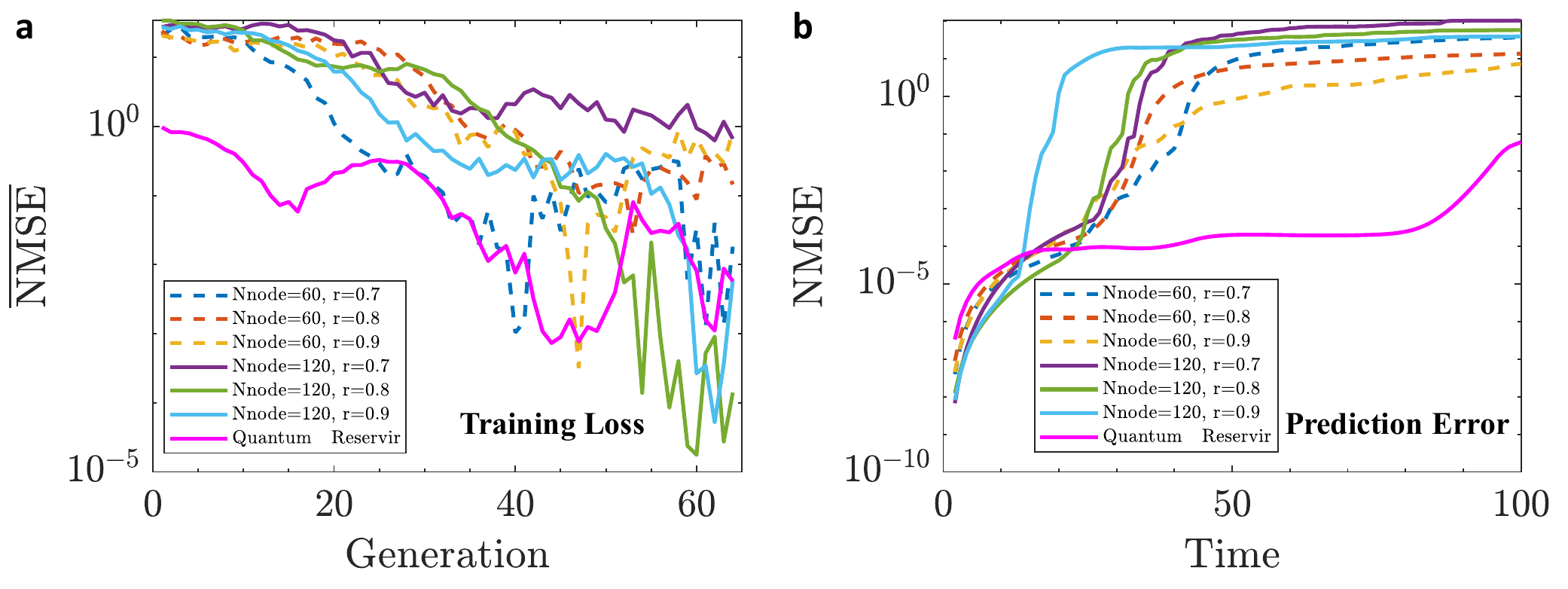}
\caption{{\bf The emergent quantum advantage with configured quantum reservoir computing}. 
{\bf a}, the average training loss during the optimization iteration by the genetic algorithm with different $N_{node} $ and ${\rm r}$. {\bf b}, the reservoir prediction accuracy on the testing dataset. Here we
choose ${\rm N_{node}}$ = 60 and 120 for ESN, and qubit number n = 6 for the quantum model. The setting of the genetic algorithm for optimization is identical for all cases, for the fairest comparison.
The examined learning tasks and the  parameter setting are the same as used in Fig.~5{\bf b} of the main text, except larger number of reservoir nodes are used here. 
}
\label{FigS2}
\end{figure}

\section{Information encoding protocol}
In this section, we demonstrate that using the Pauli-$\hat{\sigma}^X$ basis for information encoding results in significantly higher learning performance compared to using the Pauli-$\hat{\sigma}^Z$ basis. In the main text, our encoding protocol is given by $
\otimes_{j=1}^{d_{\rm in}/2} \left[ \sqrt{1-s_k(2j-1)}\ket{+}+e^{-is_k(2j)}\sqrt{s_k(2j-1)}\ket{-} \right]$, where $j$ indexes the qubits, $s_k(\ldots)$ are elements of the ${\bf s}_k$ vector, and $\ket{\pm}$ represent the eigenstates of the Pauli-$\hat{\sigma}^X$ operator. To compare the two encoding protocols, we substitute $\ket{+(-)}$ with $\ket{0(1)}$, which represent the eigenstates of the Pauli-$\hat{\sigma}^Z$ operator, while keeping everything else unchanged. We then evaluate the prediction errors of the two protocols and find that the Pauli-$\hat{\sigma}^X$ protocol produces an error that is four orders of magnitude smaller than that of the Pauli-$\hat{\sigma}^Z$ protocol for a sufficiently long period, as illustrated in Fig.~\ref{FigS3}. 
This justifies the choice of encoding basis used in the main text. 
%This is because the Pauli-$\hat{\sigma}^Z$ protocol places the input ${\bf s}_k$ in the diagonal positions, which does not generate quantum coherence, while the Pauli-$\hat{\sigma}^X$ protocol involves the off-diagonal elements and generates sufficient quantum coherence. Thus, the Pauli-$\hat{\sigma}^X$ protocol exhibits superior learning performance compared to that of the Pauli-$\hat{\sigma}^Z$ protocol.

\begin{figure}[htp]
\includegraphics[width=0.5\linewidth]{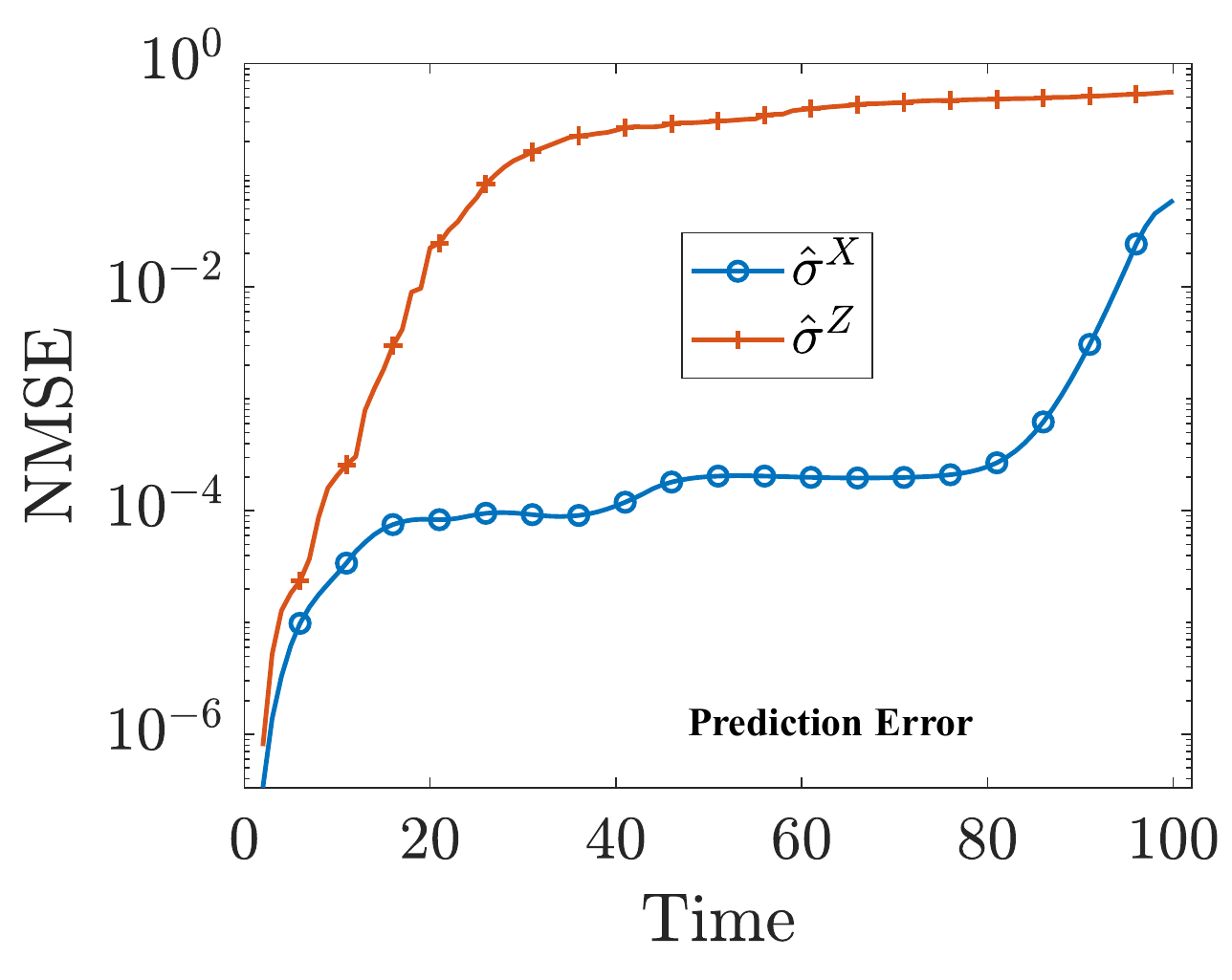}
\caption{ {\bf Comparison of quantum reservoir prediction errors between two different encoding protocols in Pauli-$\hat{\sigma}^X$ and Pauli-$\hat{\sigma}^Z$.} The entire computing process for both protocols is the same, except for the encoding basis. The examined learning tasks and parameter setting are identical to Fig.~5{\bf b} in the main text.}
\label{FigS3}
\end{figure}

\section{The effect of prior knowledge}
In the main text, the genetic algorithm (GA) is initialized with a population size of $200$. Half of the population is generated  randomly by sampling the Hamiltonian parameters in Eq.~(2),  and the other half is obtained from Refs~\onlinecite{Keisuke2017, martinez2021dynamical, xia2022reservoir, kutvonen2020optimizing}, with $25$ initial populations provided for each type of  four quantum reservoir models from the previous literature. Alternatively, we can also randomly initialize the population without invoking any prior knowledge and the results are shown in Fig.~\ref{FigS4}. Quantum reservoir computing with prior knowledge has a lower prediction error than the random initial case for an intermediate period of time, but there is no significant difference.

\begin{figure}[htp]
\includegraphics[width=0.5\linewidth]{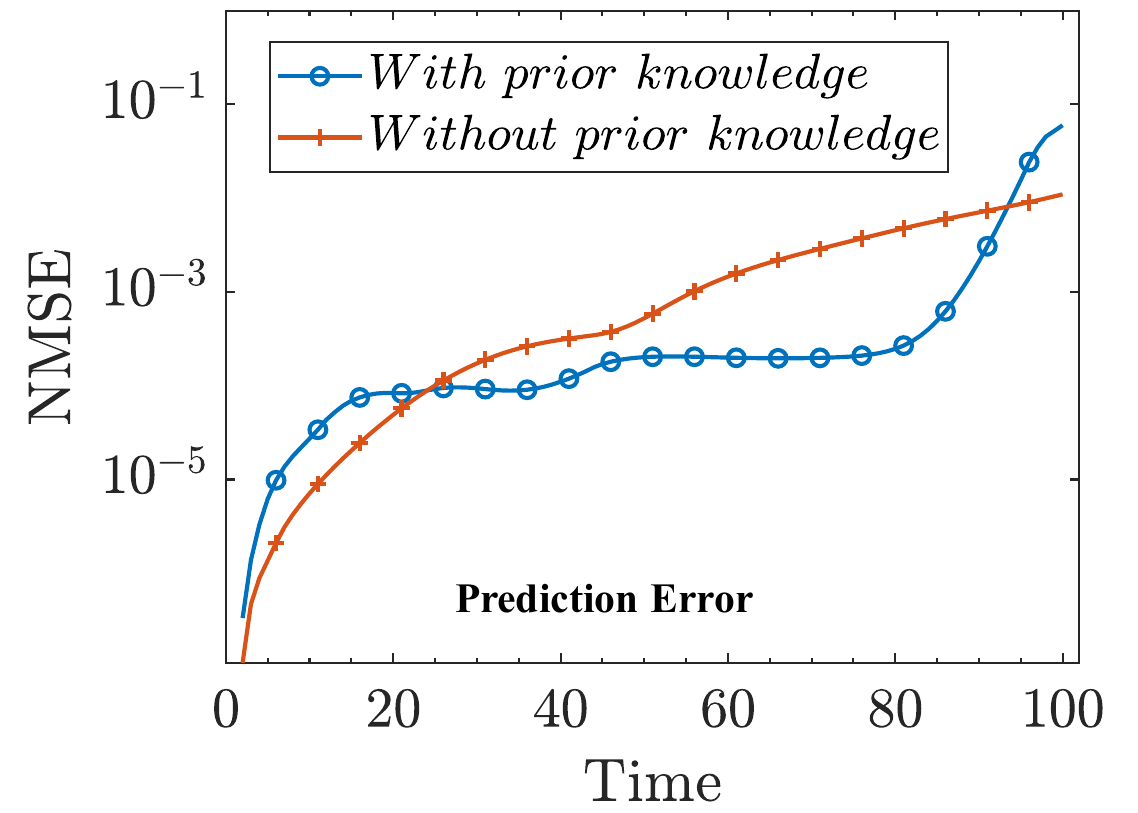}
\caption{ {\bf Comparison of quantum reservoir prediction errors between random initial populations and initial populations with prior knowledge.} The entire computing process for both cases is identical, except for the initial populations.  The examined learning tasks and the parameter setting are the same as used in Fig.~5{\bf b}. 
 }
\label{FigS4}
\end{figure}

\section{The numerical results with qubit number $n=7$}
In the main text, the number of qubits in our configured quantum reservoir is $n = 6$. In this section, we present additional numerical results for $n = 7$ qubits, which are illustrated in Fig.~\ref{FigS5}. In Fig.~\ref{FigS5}({\bf a}), we set $G_0 = 1000$, $G_1 = 6000$, and $K = 6100$. As we increase the number of qubits to $n = 7$, the prediction error decreases, particularly for longer time periods, and NMSE is approximately $10^{-5}$. We also examine longer prediction times in Fig.~\ref{FigS5}({\bf b}), where we use $G_0 = 1000$, $G_1 = 6000$, and $K = 6150$. The prediction error for the quantum reservoir with $n = 7$ is an order of magnitude lower than that of $n = 6$. The learning performance of the quantum reservoir improves as the number of qubits increases, suggesting that quantum reservoir computing has the potential to provide greater computing power with additional quantum computing resources.

\begin{figure}[htp]
\includegraphics[width=1\linewidth]{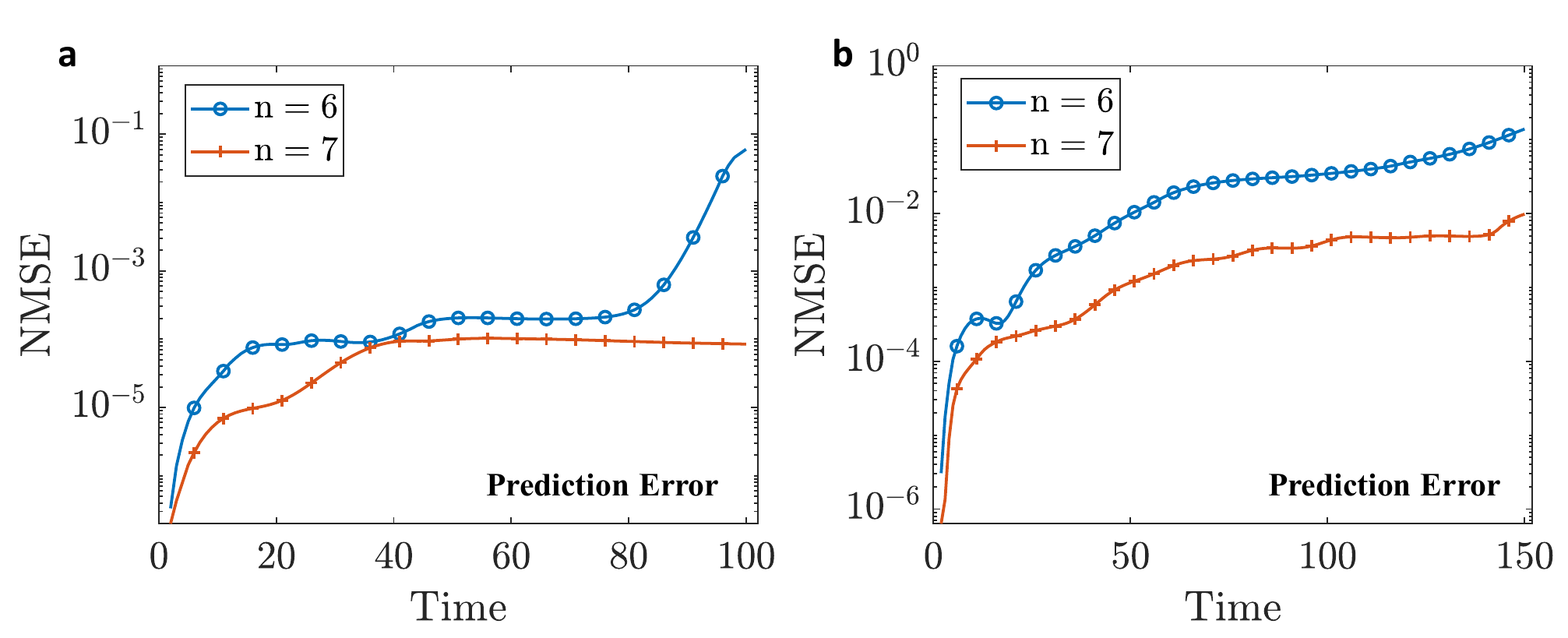}
\caption{ {\bf Comparison of quantum reservoir prediction errors between $n = 6$ and $n = 7$.} {\bf a }, we set $G_0 = 1000$, $G_1 = 6000$, and $K = 6100$. The NMSE is approximately $10^{-5}$. {\bf b }, we set $G_0 = 1000$, $G_1 = 6000$, and $K = 6150$. The remaining parameters in these two figures are chosen the same as used in Fig.~5{\bf b}. 
 }
\label{FigS5}
\end{figure}

\end{widetext}

\end{document}